\documentclass[12pt,a4paper,reqno]{article}

\usepackage{amsmath,latexsym,amssymb,amsthm}
\usepackage{graphicx}

\usepackage{amsmath}\usepackage{amssymb}
\usepackage{amsthm}

\newcommand{\be}{\begin{equation}}
\newcommand{\ee}{\end{equation}}

\renewcommand{\>}{\rangle}

\theoremstyle{definition}

\theoremstyle{remark}

\begin{document}

\title{\bf{The Future's Not Ours to See}}

\author{Anthony Sudbery$^1$\\[10pt] \small Department of Mathematics,
University of York, \\[-2pt] \small Heslington, York, England YO10 5DD\\
\small $^1$ as2@york.ac.uk}

\date{}

\maketitle

%\begin{abstract}
%
% There's going to be an abstract here which will extend over several lines and will look something like this and might be even more inane but I suspect the editor is going to want me to write one. Maybe not, since this is not a journal publication waffle waffle waffle This chapter is concerned with the limits of human \emph{knowledge} rather than understanding; in particular the limits on  our knowledge that is imposed by the nature of time. 
%
%\end{abstract}

\begin{verse}
\hspace{90pt}Que sera sera\\
\hspace{90pt}Whatever will be will be\\
\hspace{90pt}The future's not ours to see\\
\hspace{90pt}Que sera sera.
\end{verse}

So sang Doris Day in 1956, expressing a near-universal belief of humankind. You can't know the future. In this chapter I will trace the different forms of this belief, both pre-scientific and scientific, and discuss some differing kinds of scientific justification for it in physics, culminating in the form of the statement provided by the best physical theory we have found to date, namely quantum mechanics. I will argue that quantum mechanics casts doubt on the second line of the song, which suggests that even if we can't know it, there \emph{is} a definite future (and also that we can't do anything to change it -- something I will not discuss). This denial is also an ancient belief: the future is \emph{open}. If this line of the song expresses fatalism, the denial of it might be related to the existence of free will, but, again, I will not discuss this.

If it is not quite a universal belief of humankind that we \emph{can't} know the future, the universal \emph{experience} of humankind is that we \emph{don't} know the future. We don't know it, that is, in the immediate way that we know parts of the present and the past. We see some things happening in the present, we remember some things in the past, but we don't see or remember the future. But perception can be deceptive, and memory can be unreliable; this kind of direct knowledge is not certain. And there are kinds of indirect knowledge of the future which can be as certain as anything we know by direct perception or memory. I reckon I know that the sun will rise tomorrow; if I throw a stone hard at my kitchen window, I know that it will break the window. On the other hand, I did not know on Christmas Eve last year that my home town of York was going to be be hit by heavy rain on Christmas Day and nearly isolated by floods on Boxing Day.

In the ancient world, and, I think, to our childhood selves, it is events like the York floods that make us believe that we cannot know the future. I may know some things about the future, but I cannot know everything; I am sure that some things will happen tomorrow that I have no inkling of, and that I could not possibly have known about, today. In the past such events might have been attributed to the unknowable will of the gods. York was flooded because the rain god was in a bad mood, or felt like playing with us. My insurance policy refers to such catastrophes as ``acts of God"; when we feel that there is no knowing who will win an election, we say that the result is ``in the lap of the gods".

Aristotle formulated the openness of the future in the language of logic. Living in Athens at a time when invasion from the sea was always a possibility, he considered the sentence ``There will be a sea-battle tomorrow". One of the classical laws of logic is the ``law of the excluded middle" which states that every sentence is either true or false: either the sentence is true or its negation is true. But Aristotle argued that neither ``There will be a sea-battle tomorrow" nor ``There will not be a sea-battle tomorrow" is definitely true, for both possibilities lead to fatalism; if the first statement was true, for example, there would be nothing anybody could do to avert the sea-battle. Therefore these statements belong to a third logical category, neither true nor false. In modern times this conclusion has been realised in the development of many-valued logic \cite{Gottwald:manyvalued}. 

But some statements in the future tense do seem to be true; I have given the examples ``The sun will rise tomorrow" and, after I have thrown the stone, ``That window is going to break". Let's look at these more closely. In fact, none of these future statements are 100\% certain. The sun might not rise tomorrow; there might be a galactic star-trawler heading for the solar system, ready to scoop up the sun tonight and make off with it at nearly the speed of light. When I throw the stone at the window, my big brother, who is a responsible member of the family and a superb cricketer, might be coming round the corner of the house; he might see me throw the stone and catch it so as to save the window. 

We did not know that the sun would fail to make its scheduled appearance tomorrow morning; I did not know that my naughtiness would be foiled. But this lack of knowledge is not a specific consequence of the fact that we are talking about the future. If the Spaceguard programme had had a wider remit we might have seen the star-trawler coming, and then we would have known that we had seen our last sunrise; if I had known my brother's whereabouts I could have predicted his window-saving catch. In both these scenarios the lack of knowldege of the future reduces to lack of knowledge about the present.

The success of modern science gave rise to the idea that this is always true: not knowing the future can always be traced back to not knowing something about the present. As more and more phenomena came to be explained in terms of the laws of physics, so that more and more events could be explained as being caused by previous events, so confidence grew that every future event could be predicted with certainty, given enough knowledge of the present. The msot famous statement of this confidence was made by Laplace in 1814:
\begin{quote}
We may regard the present state of the universe as the effect of its past and the cause of its future. An intellect which at a certain moment would know all forces that set nature in motion, and all positions of all items of which nature is composed, if this intellect were also vast enough to submit these data to analysis, it would embrace in a single formula the movements of the greatest bodies of the universe and those of the tiniest atom; for such an intellect nothing would be uncertain and the future just like the past would be present before its eyes. (\cite{Laplace:determinism} p.4)
\end{quote}

This idea goes back to Newton, who had a dream:
\begin{quote}
       I wish we could derive the rest of the phenomena of Nature by the same kind of reasoning from mechanical principles, for I am induced by many reasons to suspect that they may all depend upon certain forces by which the particles of bodies, by some causes hitherto unknown, are either mutually impelled towards one another, and cohere in regular figures, or are repelled and recede from one another. \cite{Newton:dream}
\end{quote}
In this view, everything in the world is made up of point particles, and their behaviour is explained by the action of forces which make the particles move according to Newton's equations of motion. These completely determine the future motion of the particles if their positions and velocities are given at any one instant; the theory is \emph{deterministic}. So if we fail to know the future, that is purely because we do not know enough about the present.

For a couple of centuries Newton's dream seemed to be coming true. More and more of the physical world came under the domain of physics, as matter was analysed into molecules and atoms, and the behaviour of matter, whether chemical, biological, geological or astronomical, was explained in terms of Newtonian forces. The particles of matter that Newton dreamed of had to be supplemented by electromagnetic fields to give the full picture of what the world was made of, but the basic idea remained that they all followed deterministic laws. Capricious events like storms and floods, formerly seen as unpredictable and attributed to the whims of the gods, became susceptible to weather forecasts; and if some such events, like earthquakes, remain unpredictable, we feel sure that advancing knowledge will make them also subject to being forecast.

This scientific programme has been so successful that we have forgotten there was ever any other way to think about the future. One author writes 
\begin{quote}
In ordinary life, and in science up until the advent of quantum mechanics, \emph{all} the uncertainty that we encounter is presumed to be \ldots uncertainty arising from ignorance. \cite{Alford:spooky}
\end{quote}
We have completely forgotten what an uncertain world was inhabited by the human race before the seventeenth century, and we take Newton's dream as a natural view of waking reality.

Well, it was a nice dream. But it didn't work out that way. In the early years of the twentieth century Ernest Rutherford, investigating the recently disocvered phenomenon of radioactivity, realised that it showed random events happening at a fundamental level of matter, in the atom and its nucleus. This did not necessarily mean that Newton's dream had to be abandoned --- the nucleus is not the most fundamental level of matter, but is a complicated object made up of protons and neutrons, and maybe, if we knew exactly how these particles were situated and how they were moving, we would be able to predict when the radioactive decay of the nucleus would happen. But other, stranger, discoveries at around the same time led to the radical departure from Newtonian physics represented by quantum mechanics, which strongly reinforced the view that events at the smallest scale are indeed random, and there is no possibility of precisely knowing the future. 

The discoveries that had to be confronted by the new physics of the 1920s were twofold. On the one hand, Planck's explanation of the distribution of wavelengths in the radiation emitted by hot matter, and Einstein's explanation of the photoelectric effect, showed that energy comes in discrete packets, instead of varying continuously as it must do in Newton's mechanics and Maxwell's electromagnetic theory. On the other hand, experiments on electrons by Thomson, Davisson and Germer showed that electrons, which had been firmly established to be particles, also sometimes behaved like waves. These puzzling facts found a systematic, coherent, unified mathematical description in the theory of quantum mechanics which emerged from the work of theorists after 1926. This theory was itself so puzzling that it is not clear that it should be described as an ``explanation" of the puzzling facts it incorporated; but an essential feature of it, which seems inescapable, is that when applied to give predictions of physical effects, it yielded probabilities rather than precise numbers.

This is still not universally accepted. Some people believe that there are finer details to be discovered in the make-up of matter, which, if we knew them, would once again make it possible to predict their future behaviour precisely. This is indeed logically possible, but there would necessarily be aspects of such a theory which lead most physicists to think it highly unlikely.

The format of quantum mechanics is quite different from previous physical theories like Newtonian mechanics or electromagnetism (or both combined). These theories work with a mathematical description of the state of the world, or any part of the world; they have an equation of motion which takes such a mathematical description and tell you what it will change into after a given time. Quantum mechanics also works with a mathematical object which describes a state of the world; it is called a state vector (though it is not a vector in three dimensions like velocity), and is often denoted by the Greek letter $\psi$ or some similar symbol. But this is a different kind of mathematical description from that in mechanics or electormagnetism, which consists of a set of numbers which measure physical quantities like the velocity of a specified particle, or the electric field at a specified point of space; the quantum state vector is a more abstract object whose relation to physical quantities is indirect. From the state vector you can obtain the values of physical quantities, but only some physical quantities --- you can choose which quantities you would like to know, but you are not allowed to choose all of them. Moreover, once you have chosen which ones you would like to know, the state vector will not give you a definite answer; it will only give you probabilities for the different possible answers. This is where quantum mechanics departs from determinism. Strangely enough, in its treatment of change quantum mechanics looks like the old deterministic theories. Like them, it has an equation of motion which will tell you what a given state of the world will become after a given time; but because you can only get probabilities from this state vector, it cannot tell you what you will see after this time.

State vectors, in general, are puzzling things, and it is not at all clear how they describe physical objects. Some of them, however, do correspond to descriptions that we can understand. Among the state vectors of a cat, for example, is one describing a cat sitting and contentedly purring; there is another one descibing it lying dead, having been poisoned in a diabolical contraption devised by the physicist Erwin Schr\"odinger. But there are others, obtained mathematically by ``superposing'' these two state vectors; such a superposed state vector could be made up of a part describing the cat as live and a part describing it as dead. These are not two cats; the point of Schr\"odinger's story was that one and the same cat seems to be described as both alive and dead, and we do not understand how such states could describe anything that could arise in the real world. How can we believe this theory, generations of physicists have asked, when we never see such alive-and-dead cats?

There is an answer to this puzzle. If I were to open the box in which Schr\"odinger has prepared this poor cat, then the ordinary laws of everyday physics would ensure that if the cat was alive, I would have the image of a living cat on my retina and in my visual cortex, and the system consisting of me and the cat would end up in a state in which the cat is alive and I see a living cat. If the cat was dead, I would have the image of a dead cat, and the system consisting of me and the cat would end up in a state in which the cat is dead and I see a dead cat. It now follows, according to the laws of quantum mechanics, that if the cat is in a superposition of being alive and being dead, then the system consisting of me and the cat ends up in a superposition of the two final states described above. In symbols,
\be\label{catobserved}
|\stackrel{\centerdot\;\centerdot}{-}\>_\text{me}\big(|\text{alive}\>_\text{cat} + |\text{dead}\>_\text{cat}\big) \longrightarrow |\stackrel{\centerdot\;\centerdot}{\smile}\>_\text{me}|\text{alive}\>_\text{cat} + |\stackrel{\centerdot\;\centerdot}{\frown}\>_\text{me}|\text{dead}\>_\text{cat}.
\ee
Here a symbol like $|\Psi\>_\text{cat}$ denotes a state vector of the cat, with $\psi$ varying to describe different states (e.g.\ $\Psi = $``alive"), and similarly $|\Phi\>_\text{me}$ denotes a state vector of my body and brain, while $|\Psi\>_\text{cat}|\Phi\>_\text{me}$ denotes a state vector of the joint system of the cat and me. The plus sign between two state vectors denotes the operation of superposition (mathematically, it is very similar to the process of adding two vectors in space to give a third vector in between the first two).

Look hard at the equation \eqref{catobserved}. Nowhere in it is there a state of my brain seeing a peculiar alive-and-dead state of a cat; there are only the familiar states of seeing a live cat and seeing a dead cat. This is the answer to the question at the end of the paragraph before the last one; it follows from quantum mechanics itself that although cats have states in which they seem to be both alive and dead, we will never see a cat in such a state.

But now the combined system of me and the cat is in one of the strange superposition states introduced by quantum mechanics. It is called an \emph{entangled} state of me and the cat. How are we to understand it? We can understand the states represented by $|\stackrel{\centerdot\;\centerdot}{\smile}\>_\text{me}|\text{alive}\>_\text{cat}$ and $|\stackrel{\centerdot\;\centerdot}{\frown}\>_\text{me}|\text{dead}\>_\text{cat}$ individually --- the cat is alive in one of them, and dead in the other, and I have the corresponding visual experience --- but what does this mathematical sum, this superposition, mean? Maybe the mathematical sign $+$ just means ``or"; that would make sense. But unfortunately this meaning, if applied to the states of an electon, is not compatible with the facts of interference observed in the experiments that show the electron behaving like a wave (\cite{QMPN} p. 22--24, 207). Some people think that this $+$ should be understood as ``and": when the cat and I are in the state \eqref{catobserved}, there is a world in which the cat has died and I see a dead cat, and another world in which the cat is still alive and I see a living cat. Others do not find this a helpful picture (\cite{ManyWorlds?}, \cite{QMPN} p. 221). Let us just take it as a true description of the cat and me, whose meaning is problematic.

Now let us broaden our horizon and consider the whole universe, which contains each one of us considered as a sentient, observing physical system. According to quantum mechanics this has a description by a state vector like \eqref{catobserved}, with me replaced by any sentient obsever and the cat replaced by the rest of the universe. A sentient system, like you or me, has a large number of possible experiences, each of which occurs physically in certain states of the sentient system. These are described by state vectors of the understandable kind like  $|\stackrel{\centerdot\;\centerdot}{\smile}\>_\text{me}$ and $|\stackrel{\centerdot\;\centerdot}{\frown}\>_\text{me}$. The whole universe can then be described by a state vector generalising \eqref{catobserved}, in which the states of the cat are replaced by states of the rest of the universe which go with the possible experiences of the observer. For those who like equations, the state vector of the universe is of the form
\be\label{universe}
|\Psi(t)\> = \sum_n|\eta_n\>|\Phi_n(t)\>
\ee
where $|\Psi(t)\>$ is the state vector of the whole universe at time $t$, $|\eta_n\>$ is a state vector of the observer in which they are having the experience $\eta_n$, and $|\Phi_n(t)\>$ is the corresponding state of the rest of the universe.)

Saying that this is the truth about the universe seems to conflict with my knowledge of what I see. Let us suppose that the cat survived when I did Schrodinger's experiment. Then I know that my state is $|\stackrel{\centerdot\;\centerdot}{\smile}\>_\text{me}$ and therefore the cat's state is $|\text{alive}\>_\text{cat}$. The other part of \eqref{catobserved} is not part of the truth; it describes something that might have happened but didn't. In the general case of the whole universe, I know that I have just one of the experiences $\eta_n$ and therefore that the state of me and the rest of the universe is just one of the terms $|\eta_n\>|\Phi_n(t)\>$ and not the whole of \eqref{universe}. But this contradicts what was asserted in the previous paragraph. Which of these is the truth?

This contradiction is of the same type as many familiar contradictions between objective and subjective statements. It can be resolved in the way put forward by Thomas Nagel \cite{Nagel:subjobj,Nagel:nowhere}: we must recognise that there are two positions from which we can make statements of fact or value, and statements made in these two contexts are not commensurable. In the \emph{external} context (the God's-eye view, or the ``view from nowhere") we step outside our own particular situation and talk about the whole universe. In the \emph{internal} context (the view from \emph{now here}), we make statements as physical objects inside the universe. Thus in the external view, the state vector $|\Psi(t)\>$ is the whole truth about the universe; the components describing my different possible experiences, and the corresponding states of the rest of the universe, are (unequal) parts of this truth. But in the internal view, from the perspective of some particular experience $|\eta_n\>$ which I know I am having, that experience, together with the corresponding state of the rest of the universe, is the actual truth. I may know what the other components are, because I can calculate $|\Psi(t)\>$ from the Schr\"odinger equation; but these other components, for me, represent things that \emph{might} have happened but \emph{didn't}. 

We can now look at what quantum mechanics tells us about the future. As we should now expect, there are two answers, one for each of the two perspectives. From the external perspective, the universe at any one time is described by a universal state vector, and state vectors at different times are related by the Schr\"odinger equation. Given the state vector at the present time, the Schr\"odinger equation delivers a unique state vector at any future time: the theory is completely deterministic, in complete accord with Pascal's world-view (in a quantum version).

From the internal perspective, however, things are completely different. We now have to specify a particular observer (who has been me in the above discussion, but it could have been you or anyone else, or indeed the whole human race taken together), with respect to which we can carve up the universal state vector as in \eqref{universe}; and we have to specify a particular experience state of that observer, say $|\eta_0\>$. From the perspective of that experience, it is by definition true that the observer has the definite experience $\eta_0$, and it follows from \eqref{universe} that the universe is in a product state $|\eta_0\>|\Phi_0(t)\>$. The Schr\"odinger equation applies, as before. But with this initial state the Schr\"odinger equation will yield a state of the universe at a future time which, in general, is not a product state; in particular, in situations like that of Schr\"odinger's cat in which quantum effects are amplified to a macroscopic level at which they can be seen by our observer, the initial product state will develop into a superposition like the right-hand side of \eqref{catobserved}. So at the future time there will be a universal state vector of the form \eqref{universe}, in which most terms will be zero but more than one experience state of the observer occurs with non-zero coefficient. (A ``coefficient" here is a state vector of the rest of the universe, the magnitude being significant.) What is the present observer to make of this? The experience states in the non-zero components must describe possible experiences of the observer at the future time $t$. No one of them is singled out as being what the observer actually will experience at that time. Some of them, though, loom larger than others, accordng to the magnitudes of their coefficients in the universal state vector. 

I found this rather startling when I first encountered it. I was used to thinking that there is something awaiting me in the future, even if I cannot know what it is, and even if there is no law of nature which determines what it is. Whatever will be will be, indeed. But Aristotle already saw that this is wrong. A statement about the future, in general, cannot be actually true, even when we are careful to distinguish being true from our knowing that it is true. But we can say more than that. 

Aristotle pointed out that although no one statement about the future is actually true, some of them are more \emph{likely} than others. Similarly, the universal state vector at the time $t$ contains more information, for me, than my possible experiences at time $t$. These experience states, occurring as components of the universal state vector, contribute to it to different extents, measured by the magnitudes of their coefficients. Such magnitudes are usually used in quantum mechanics to calculate \emph{probabilities}. So we can understand the future universal state as giving information, not only about what experiences are possible for me at that future time, but also about how probable each experience is. 

The nature of proabability is a long-standing philosophical problem \cite{Gillies}, to which scientists also need an answer. Many scientists take the view that the probability of an event only makes sense when there are many repetitions of the circumstatnces in which the event might occur, and we count up the number of times that it does occur; they hold that the probability of a single, unrepeated event does not make sense. But what we have just outlined does seem to be a calculation of a single event at a time which will only come once. In everyday life we often talk about the probability that something will happen on just one occasion --- that it will rain tomorrow, or that a particular horse will win a race, or that there will be a sea-battle. A standard view of such single-event probability is that it refers to the strength of the belief of the person who is asserting the probability, and can be measured by the betting odds they are prepared to offer on the event happening. But the probability described in the previous paragraph is an objective fact about the universe. It has nothing to do with the beliefs of an individual, not even the individual whose experiences are in question; that individual is being told a fact about their future experiences, whether they believe it or not. What does this probability mean?

The probability that I will have a particular experience tomorrow --- that I will see the cat alive, for example --- arose in the context that there is no fact about what experience I will have tomorrow; the statement ``I will see a living cat" is neither true nor false. In logical terms, its truth value is neither 1 nor 0. But if the probability of this experience is close to 1, the statement is nearly true; if it close to 0, the statement is nearly false. This suggests that the truth value of the statement should be identified with its probability, and that this tells us what the probability of a single event means. \emph{The probability of a future event is the truth value of the future-tense proposition that that event will happen}. This view of probability, and the associated many-valued logic of tensed propositions, has been explored in \cite{logicfuture}.

It has now become clear that the universal state vector plays very different roles in the two perspectives, external and internal. From the external perspective, it is a full description of reality; it tells how the universe is constituted at a particular time. This complete reality can be analysed with respect to any sentient system as in \eqref{universe}, yielding a number of components, attached to different experiences of the chosen sentient system, which are all parts of the universal reality. From the internal perspective of this system, however, reality consists of just one of these experiences; the component attached to this experience is the complete truth about the universe for the sentient system. All the other non-zero components are things that \emph{might} have happened, but \emph{didn't}. The role of the universal state vector at a later time, in this perspective, is not to describe how the universe \emph{will be} at that time, but to specify how the present state of the universe \emph{might change} between now and then. It gives a list of possibilities at that later time, with a probability for each of them that it will become the truth.

Human knowledge of the future, therefore, is limited in a fundamental way. It is not that there are true facts but the knowledge of them is not accessible to us; there are no facts out there, and there is simply no certain knowledge to be had. Nevertheless, there are facts with partial degrees of truth, and our knowledge of them is itself partial. Our best knowledge of the future can only be probable.

%\bibliography{quantum}
%\bibliographystyle{plain}

\end{document}